\documentstyle[12pt,aps,preprint]{revtex}

\def\be{\begin{equation}}
\def\ee{\end{equation}}

\begin{document}

\begin{center}
\baselineskip=15pt plus 1pt minus 1pt
{\Large \bf Quantal Modifications to the Wheeler DeWitt Equation} \\[3ex]
{\large R.~Jackiw} \\
{\it Center for Theoretical Physics, \\
Laboratory for Nuclear Science \\
and Department of Physics, \\
Massachusetts Institute of Technology, \\
Cambridge, MA ~02139--4307}
\end{center}

\vspace*{-.5in}

\begin{abstract}
\vspace*{-.45in}

\baselineskip=13pt
Commutator anomalies obstruct solving the Wheeler-DeWitt constraint
equation in Dirac quantization of quantum gravity-matter theory.  When
the obstruction is removed, there result quantal modifications to the
constraints.  The same classical theory gives rise to different quantum
theories when different procedures for overcoming anomalies are
implemented.
\end{abstract}

\footnote%
{%
\noindent MIT-CTP-2442
\hfil gr-qc/xxxxxxx
\hfil CAM'95, Qu\'ebec, Canada, June 1995
\break}

\baselineskip=14pt plus 1pt minus 1pt
\parskip=1ex

In a canonical, Hamiltonian approach to quantizing a theory with local
symmetry --- a theory that is invariant against transformations whose
parameters are arbitrary functions on space-time ---  there occur
constraints, which are imposed on physical states.   Typically these
constraints correspond to time components of the Euler-Lagrange
equations, and familiar examples arise in gauge theories.   The time
component of the gauge field equation is the Gauss law.
\be
G_a \equiv {\bf D} \cdot {\bf E}^a - \rho^a = 0
\ee
Here ${\bf E}^a$ is the (non-Abelian) electric field, $\rho^a$ the
matter charge density, and ${\bf D}$ denotes the gauge-covariant
derivative.  When expressed in terms of canonical variables, $G_a$ does
{\it not\/} involve time-derivatives --- it depends on canonical
coordinates and momenta, which we denote collectively by the symbols $X$
and $P$ respectively  ($X$ and $P$ are fields defined at fixed time)
$: ~ G_a = G_a (X,P)$.   Thus in a Schr\"odinger representation for the
theory, the Gauss law condition on physical states
\be
G_a (X,P) | \, \psi \, \rangle = 0
\ee
corresponds to a (functional)  differential equation that the state
functional $\Psi(X)$ must satisfy.
\be
G_a \left( X, {1\over i} {\delta \over \delta X} \right)
\, \Psi(X) = 0
\label{eq:3}
\ee
In fact, Eq.~(\ref{eq:3}) represents an infinite number of equations,
one for each spatial point ${\bf r}$, since $G_a$ is also the generator
of the local symmetry:
$G_a = G_a({\bf r})$.
Consequently, questions of consistency (integrability) arise, and these
may be examined by considering the commutator of two constraints.
Precisely because the $G_a$ generate the symmetry transformation,
one expects their commutator to follow the Lie algebra with structure
constants $f_{abc}$.
\be
\left[ G_a ({\bf r}) , G_b (\tilde{\bf r}) \right]
= i \, f_{abc} \, G_c ({\bf r}) \, \delta({\bf r} - \tilde{\bf r})
\label{eq:4}
\ee
If (\ref{eq:4}) holds, the constraints are consistent --- they are first
class --- and the constraint equations are integrable, at least locally.

However, it is by now well-known that Eq.~(\ref{eq:4}), which {\it
does\/} hold classically with Poisson bracketing, may acquire a quantal
anomaly.  Indeed when the matter charge density is constructed from
fermions of a definite chirality, the Gauss law algebra is modified by
an extension --- a Schwinger term --- the constraint equations become
second-class and Eq.~(\ref{eq:3}) is inconsistent and cannot be solved.
We call such gauge theories ``anomalous.''

This does not mean that a quantum theory cannot be constructed from an
anomalous gauge theory.  One can adopt various strategies for overcoming
the obstruction, but these represent modifications of the original
model.  Moreover, the resulting quantum theory possesses physical
content that is very far removed from what one might infer by studying
the classical model.  All this is explicitly illustrated by the
anomalous chiral Schwinger model, whose Gauss law is obstructed, while a
successful construction of the quantum theory leads to massive
excitations, which cannot be anticipated from the un-quantized equations
[1].

With these facts in mind, we turn now to gravity theory, which obviously
is invariant against local transformations that redefine coordinates of
space-time.

Indeed over the years there have been many attempts to describe gravity
in terms of a gauge theory.  That program is entirely successful in
three- and two-dimensional space-time, where gravitational models are
formulated in terms of Einstein--Cartan variables (spin-connection, {\it
Vielbein}) as non-Abelian gauge theories, based not on the Yang-Mills
paradigm, but rather on Chern-Simons and B-F structures.

But even remaining with the conventional metric-based formulation, it
is recognized that the time components of Einstein's equation comprise
the constraints.
\be
{1\over   8\pi G}  \left( R_\nu^{\,0} -
{\textstyle{1\over2}} \delta_\nu^{\,0} R \right)
- T_\nu^{\,0} = 0
\label{eq:5}
\ee
The gravitational  part is the time component of the Einstein tensor
$R^\mu_{\nu} - {1\over2} \delta^\mu_{\nu} R$;
weighted by Newton's constant $G$, this equals the time component of the
matter energy-momentum tensor, $T^{\mu}_{\nu}$.   In the quantized
theory, the collection of canonical operators on the left side in
(\ref{eq:5}) annihilates physical states.  The resulting equations may be
presented as
\begin{eqnarray}
{\cal E} \, | \, \psi \, \rangle &=& 0 ~~,
\label{eq:6} \\
{\cal P}_i \, | \, \psi \, \rangle &=& 0 ~~,
\label{eq:7}
\end{eqnarray}
where ${\cal E}$ is the energy constraint
\be
{\cal E} = {\cal E}^{\rm\,gravity} + {\cal E}^{\rm\,matter} ~~,
\ee
and ${\cal P}_i$ is the momentum constraint.
\be
{\cal P}_i = {\cal P}_i^{\rm\,gravity} + {\cal P}_i^{\rm\,matter}
\ee
Taking for definiteness matter to be described by a massless, spinless
field $\varphi$, with canonical momentum $\Pi$, we have
\begin{eqnarray}
{\cal E}^{\rm\,matter} &=&
{\textstyle{1\over2}} \left( \Pi^2 + \gamma \,
\gamma^{ij} \, \partial_i \, \varphi \, \partial_j \, \varphi \right) \\
{\cal P}_i^{\rm\,matter} &=& \partial_i \, \varphi \, \Pi
\end{eqnarray}
Here $\gamma_{ij}$ is the spatial metric tensor;
$\gamma$, its determinant;
$\gamma^{ij}$, its inverse.

The momentum constraint in Eq.~(\ref{eq:7}) is easy to unravel.
In a Schr\"odinger representation, it requires that
$\Psi(\gamma_{ij}, \varphi)$
be a functional of the canonical field variables
$\gamma_{ij},\varphi$
that is invariant against reparameterization of the spatial
coordinates and such functionals are easy to construct.

Of course it is (\ref{eq:6}), the Wheeler-DeWitt equation, that is
highly non-trivial and once again one asks about its consistency.  If
the commutators of ${\cal E}$ with ${\cal P}$ follow their Poisson
brackets one would expect
that the following algebra holds.
\begin{mathletters}%
\begin{eqnarray}%
\left[ {\cal P}_i ({\bf r}), {\cal P}_j (\tilde{\bf r}) \right]
&=& i {\cal P}_j ({\bf r}) \, \partial_i \, \delta({\bf r} - \tilde{\bf r})
+ i {\cal P}_i (\tilde{\bf r}) \, \partial_j \, \delta({\bf r} - \tilde{\bf r})
\label{eq:12a} \\
\left[ {\cal E} ({\bf r}), {\cal E} (\tilde{\bf r}) \right]
&=& i \left( {\cal P}^i ({\bf r}) + {\cal P}^i (\tilde{\bf r}) \right)
\, \partial_i \, \delta({\bf r} - \tilde{\bf r})
\label{eq:12b} \\
\left[ {\cal E} ({\bf r}), {\cal P}_i (\tilde{\bf r}) \right]
&=& i \left( {\cal E} ({\bf r}) + {\cal E} (\tilde{\bf r}) \right)
\, \partial_i \, \delta({\bf r} - \tilde{\bf r})
\label{eq:12c}
\end{eqnarray}%
\end{mathletters}%
Here ${\cal P}^i \equiv \gamma \, \gamma^{ij} \, {\cal P}_j$.
If true, Eqs.~(12) would demonstrate the consistency of the
constraints, since they appear first-class.  Unfortunately, establishing
(12) in the quantized theory is highly problematical.  First
of all there is the issue of operator ordering in the gravitational
portion of ${\cal E}$ and ${\cal P}$.  Much has been said about this,
and I shall not address that difficulty here.

The problem that I want to call attention to is the very likely
occurrence of an extension in the $[{\cal E}, {\cal P}_i]$ commutator
(\ref{eq:12c}).  We know that in flat space, the commutator between the
matter energy and momentum densities possesses a triple derivative
Schwinger term [2].  There does not appear any known mechanism
arising from the gravity variables
that would effect a cancellation of this obstruction.

A definite resolution of this question in the full quantum theory is out
of reach at the present time.  Non-canonical Schwinger terms can be
determined only after a clear understanding of the singularities in the
quantum field theory and the nature of its Hilbert space are in hand,
and this is obviously lacking for four-dimensional quantum gravity.

Faced with the impasse, we turn to a gravitational model in
two-dimensional space-time
--- a {\it lineal\/} gravity theory ---
where the calculation can be carried to a
definite conclusion: an obstruction does exist and the model is
anomalous.  Various mechanisms are available to overcome the anomaly,
but the resulting various quantum theories
are inequivalent and bear little resemblance to the classical model.

In two dimensions, Einstein's equation is vacuous because
$R_\nu^\mu = {1\over2} \delta^\mu_\nu R$;
therefore gravitational dynamics has to be invented afresh.   The models
that have been studied recently posit local dynamics for the ``gravity''
sector, which involves as variables the metric tensor and an additional
world scalar (``dilaton'' or Lagrange multiplier)  field.  Such
``scalar-tensor'' theories, introduced a decade ago [3], are obtained by
dimensional reduction from higher-dimensional Einstein theory [3,4].
They should be contrasted with models where quantum fluctuations of
matter variables induce gravitational dynamics [5], which therefore are
non-local and do not appear to offer any insight into the questions
posed by the physical, four-dimensional theory.

The model we study is the so-called ``string-inspired dilaton gravity''
-- CGHS theory [6].   The gravitational action involves the metric
tensor $g_{\mu\nu}$, the dilaton field $\phi$, and a cosmological
constant $\lambda$.  The matter action describes the coupling of a
massless, spinless field $\varphi$.
\begin{eqnarray}
I_{\rm\,gravity} &=& \int d^2x \, \sqrt{-g} \, e^{-2\phi}
\left( R + 4 g^{\mu\nu} \partial_\mu \phi \partial_\nu \phi - \lambda
\right) \label{eq:13} \\
I_{\rm\,matter} &=& {\textstyle{1\over2}} \int d^2 x \,
\sqrt{-g} \, g^{\mu\nu} \, \partial_\mu \varphi \partial_\nu \varphi
\label{eq:14}
\end{eqnarray}
The total action is the sum of (\ref{eq:13}) and (\ref{eq:14}),
weighted by ``Newton's'' constant $G$:
\be
I = {1\over 4\pi G} ~ I_{\rm\,gravity} + I_{\rm\,matter}
\label{eq:15}
\ee

In fact this theory can be given a gauge-theoretical ``B-F''
description
based on the
centrally extended
Poincar\'e group
in (1+1) dimensions [7].
This formulation aided us immeasurably    in the subsequent
analysis/transformations.
However, I shall not discuss this here,
because in retrospect it proved possible
to carry the analysis forward within the
metric formulation (13)--(15).

After a remarkable sequence of redefinitions and canonical
transformations on the dynamical variables in (13)--(15),
one can present $I$ in terms
of a first-order Lagrange density ${\cal L}$ that is a sum of
quadratic terms.
\begin{eqnarray}
{\cal L} &=& \pi_a \dot{r}^a + \Pi \dot{\varphi}
- \alpha {\cal E} - \beta {\cal P}
\label{eq:16} \\
{\cal E} &=& - {\textstyle{1\over2}}
\left(
{\textstyle{1\over \Lambda}} \pi^a \pi_a + \Lambda {r^a}' {r_a}' \right)
+ {\textstyle{1\over2}} \left( \Pi^2 + {\varphi'}^2 \right)
\label{eq:17} \\
{\cal P} &=& - {r^a}' \pi_a  - \varphi' \Pi
\label{eq:18}
\end{eqnarray}
I shall not derive this, but merely explain it.
The index ``a'' runs over flat 2-dimensional $(t,\sigma)$ space, with
signature $(1,-1)$.  Dot (dash) signify differentiation with respect to
time $t$ (space $\sigma$).  The four variables
$\left\{ r^a, \alpha, \beta \right\}$ correspond to the four
gravitational variables $(g_{\mu\nu}, \phi)$, where only $r^a$ is
dynamical with canonically  conjugate momentum $\pi_a$, while $\alpha$
and $\beta$ act as Lagrange multipliers.  Notice that regardless of the
sign $\Lambda \equiv \lambda / 8 \pi G$, the gravitational
contribution to ${\cal E}$, is quadratic with indefinite sign.
\begin{mathletters}%
\begin{eqnarray}
{\cal E}^{\rm\,gravity} &=&
- {\textstyle{1\over2}} \left(
{\textstyle{1\over\Lambda}} \pi^a \pi_a +
\Lambda {r^a}' {r_a}' \right) \nonumber \\
&=& - {\textstyle{1\over2}} \left(
{\textstyle{1\over\Lambda}} (\pi_0)^2
- {\textstyle{1\over\Lambda}} (\pi_1)^2
+ \Lambda ({r^0}')^2
- \Lambda ({r^1}')^2 \right) \nonumber\\
&=& - {\cal E}_0 + {\cal E}_1 \\
{\cal E}_0 &=& {\textstyle{1\over2}} \left(
{\textstyle{1\over\Lambda}}
(\pi_0)^{2} + \Lambda ({r^0}')^2 \right) \\
{\cal E}_1 &=& {\textstyle{1\over 2}}
\left( {\textstyle{1\over\Lambda}}
(\pi_1)^{2} + \Lambda ({r^1}')^2 \right)
\end{eqnarray}
\end{mathletters}%
On the other hand, the gravitational contribution to the momentum does
not show alteration of sign.
\begin{mathletters}%
\begin{eqnarray}
{\cal P}^{\rm\,gravity} &=& - {r^a}' \pi_a \nonumber\\
                        &=& - {r^0}' \pi_0  - {r^1}' \pi_1 \nonumber\\
                        &=& {\cal P}_0 + {\cal P}_1 \\
{\cal P}_0 &=& - {r^0}' \pi_0 \\
{\cal P}_1 &=& - {r^1}' \pi_1
\end{eqnarray}
\end{mathletters}%
One may understand the relative negative sign        between the two
gravitational    contributors $(a=0,1)$ as follows.   Pure   metric
gravity   in two space-time dimensions is described by three functions
collected   in $g_{\mu\nu}$.  Diffeomorphism invariance involves 2
functions, which reduce the number of variables by $2\times2$,
{\it i.e.\/} pure gravity has $3-4=-1$ degrees of freedom.  Adding the
dilaton $\phi$ gives a net number of $-1+1=0$, as in our final
gravitational Lagrangian.

The matter contribution    is the conventional expression for massless
and spinless fields:
\begin{eqnarray}
{\cal E}^{\rm\,matter} &=& {\textstyle{1\over2}} (\Pi^2 + {\varphi'}^2) \\
{\cal P}^{\rm\,matter} &=& - \varphi' \, \Pi
\end{eqnarray}
It is with the formulation in Eqs.~(16)--(22) of the theory (13)--(15)
that we embark upon the various quantization procedures.

The transformed theory appears very simple:  there are three independent
dynamical fields
$\left\{ r^a, \varphi \right\}$
and together with the canonical momenta
$\left\{ \pi_a, \Pi \right\}$
they lead to a quadratic Hamiltonian, which has
no interaction terms among the three.  Similarly, the momentum comprises
non-interacting terms.  However, there remains a subtle ``correlation
interaction'' as a consequence of the constraint that ${\cal E}$ and
${\cal P}$ annihiliate physical states, as follows from varying the
Lagrange multipliers $\alpha$ and $\beta$ in (\ref{eq:16})
\begin{eqnarray}
{\cal E} \, | \, \psi \, \rangle  &=& 0 ~~, \label{eq:23} \\
{\cal P} \, | \, \psi \, \rangle  &=& 0 ~~. \label{eq:24}
\end{eqnarray}
Thus, even though ${\cal E}$ and ${\cal P}$ each are sums of
non-interacting variables, the physical states
$| \, \psi \, \rangle$ are not      direct products of states for the
separate degrees of freedom.  Note that Eqs.~(\ref{eq:23}), (\ref{eq:24})
comprise the entire physical content of the theory.   There is no need
for any further ``gauge fixing'' or ``ghost'' variables --- this is the
advantage of the Hamiltonian formalism.

As in four dimensions, the momentum constraint
(24)
enforces invariance
of the state functional $\Psi(r^a, \varphi)$ against spatial coordinate
transformations, while the energy constraint (23)
--- the Wheeler-DeWitt
equation in the present lineal gravity context --- is highly non-trivial.

Once again one looks to the algebra
of the constraints
to check consistency.  The reduction
of (12) to one spatial dimension leaves (after the identification
${\cal P}_i \to - {\cal P}, \gamma \gamma^{ij} \to 1$)
\begin{mathletters}%
\begin{eqnarray}
i [ {\cal P}(\sigma), {\cal P}(\tilde{\sigma}) ] &=&
\left( {\cal P}(\sigma) + {\cal P}(\tilde{\sigma}) \right)
\, \delta'(\sigma - \tilde{\sigma}) \\
i [ {\cal E}(\sigma), {\cal E}(\tilde{\sigma}) ] &=&
\left( {\cal P}(\sigma) + {\cal P}(\tilde{\sigma}) \right)
\, \delta'(\sigma - \tilde{\sigma}) \\
i [ {\cal E}(\sigma), {\cal P}(\tilde{\sigma}) ] &=&
\left( {\cal E}(\sigma) + {\cal E}(\tilde{\sigma}) \right)
\, \delta'(\sigma - \tilde{\sigma})
- {c \over 12\pi} \delta''' (\sigma - \tilde{\sigma})
\end{eqnarray}
\end{mathletters}%
where we have allowed for a possible central extension of strength $c$,
and it remains to calculate this quantity.

The gained advantage in two dimensional space-time is that all operators
are quadratic, see (19)-(22); the singularity structure may be assessed
and $c$ computed; obviously it is composed of independent contributions.
\be
c = c^{\rm\,gravity} + c^{\rm\,matter} ~~,~~~~~~
c^{\rm\,gravity} = c_0 + c_1
\ee
Surprisingly, however,
there is more than one way of handling infinities and more than one
answer for $c$ can be gotten.  This reflects the fact, already known to
Jordan in the 1930s [8], that an anomalous Schwinger term depends
on how the vacuum is defined.

In the present context, there is no argument about
$c^{\rm\,matter}$, the answer is
\be
c^{\rm\,matter} = 1
\ee
The same holds for the positively signed gravity variable
(assume $\Lambda > 0$, so that $a_1$ enters positively).
\be
c_1 = 1
\ee

But the negatively signed gravitational variable can be treated   in
more than one way, giving different answers for $c_0$.  The different
approaches may be named
``Schr\"odinger representation quantum field theory''
and ``BRST string/conformal field theory,''
and the variety arises owing to the various ways
one can quantize a theory with a negative kinetic term, like the $r^0$
gravitational variable.  (This variety is analogous to what is seen in
Gupta-Bleuler quantization of electrodynamics: the time component
potential $A_0$ enters with negative kinetic term.)

In the Schr\"odinger representation quantum field theory approach one
maintains positive norm states in a Hilbert space, and finds $c_0 = -1$,
$c^{\rm gravity} = c_0 + c_1 = 0$, $c = c^{\rm gravity} + c^{\rm matter}
= 1$.  Thus pure gravity has no obstructions, only matter provides the
obstruction.  Consequently the constraints of pure gravity can be
solved, indeed explicit formulas have been gotten by many people [9].
In our formalism, according to (19) and (20) the constraints read
\begin{eqnarray}
{\cal E}^{\rm gravity} \big| \psi \big>_{\rm gravity} & \sim & {1 \over 2}
     \left({1 \over \Lambda} {\delta^2 \over \delta r^a \delta r_a} -
          \Lambda {r^a}' {r_a}' \right) \Psi_{\rm gravity} (r^a) = 0  \\
{\cal P}^{\rm gravity} \big| \psi \big>_{\rm gravity} & \sim & i {r^a}'
     {\delta \over \delta r^a} \Psi_{\rm gravity} (r^a) = 0
\end{eqnarray}
with two solutions
\begin{mathletters}
\begin{equation}
\Psi_{\rm gravity} (r^a) = {\rm exp} \pm i {\Lambda \over 2}
     \int d \sigma \epsilon_{ab} r^a {r^b}'
\end{equation}
This may also be presented by an action of a definite operator on the Fock
vacuum state $\big| 0 \big>$,
\begin{equation}
\Psi_{\rm gravity} (r^a) \propto \left[ {\rm exp} \pm \int dk \,
  {a_0}^{\!\dagger} (k) \, \epsilon (k) \,
  {a_1}^{\!\dagger} (-k) \right] \big| \, 0 \, \big>.
\end{equation}
with $  {a_a}^{\!\dagger} (k)$ creating field oscillations of definite
momentum.
\begin{equation}
{a_a}^{\!\dagger} (k)  = {-i \over \sqrt{4 \pi \Lambda |k|}}
\int d \sigma \, e^{i k \sigma} \, \pi_a (\sigma)
+ \sqrt{ {\Lambda |k| \over 4\pi} }
\int d \sigma \, e^{i k \sigma} \, r^a (\sigma)
\end{equation}
\end{mathletters}%
As expected, the state functional is invariant against
spatial coordinate redefinition, $\sigma \to \tilde{\sigma}(\sigma)$;
this is best seen by recognizing that integrand in the exponent of (31a)
is a 1-form:
$d\sigma \, \epsilon_{ab} \, r^a \, {r^b}' = \epsilon_{ab} \, r^a \, dr^b$.
[It is important that {\it two\/} fields, $r^0$ and $r^1$, are in
play.  One cannot construct an invariant functional out of just one field.]

Although this state is here presented
for a gravity model
in the Schr\"odinger
representation field theory context, it is also of interest to
practitioners of conformal field theory and string theory.  The algebra
(25), especially when written in decoupled form,
\begin{equation}
\Theta_{\pm} = {1 \over 2} ({\cal E} \mp P)
\end{equation}
\vspace*{-0.3truein}
\begin{mathletters}
\begin{eqnarray}
\left[ \Theta_{\pm} (\sigma), \Theta_{\pm} (\tilde{\sigma}) \right] & = &
  \pm i \left( \vphantom{1\over2}
\Theta_{\pm} (\sigma) + \Theta_{\pm} (\tilde{\sigma}) \right)
  \delta' (\sigma - \tilde{\sigma}) \mp {ic \over 24\pi} \delta'''
      (\sigma - \tilde{\sigma})\\
\left[ \Theta_{\pm} (\sigma), \Theta_{\mp} (\tilde{\sigma}) \right] & = & 0
\end{eqnarray}
\end{mathletters}%
is recognized as the position-space version of the Virasoro algebra and
the Schwinger term is just the Virasoro anomaly.  Usually one does not
find a field theoretic non-ghost realization {\it without\/} the Virasoro
center; yet the CGHS model, without matter provides an explicit example.
Usually one does not expect that {\it all\/} the Virasoro generators
annihilate a state, but in fact our states (31) enjoy that property.

Once matter is added, a center appears, $c=1$, and the theory becomes
anomalous.  In the same Schr\"odinger representation approach used
above, one strategy is the following
modification of a method due to Kucha\v{r}
[7,10].  The Lagrange density (16)
is presented in terms of decoupled constraints.
\begin{mathletters}
\begin{equation}
{\cal L} = \pi_a \dot{r}^a + \Pi \dot{\varphi} - \lambda^+ \Theta_+ - \lambda^-
\Theta_-
\end{equation}
\begin{equation}
\lambda^{\pm} = \alpha \pm \beta
\end{equation}
\end{mathletters}%
Then the gravity variables $\{ \pi_a, r^a \}$
are transformed by a linear canonical transformation to a new set
$\{ P_\pm, X^{\pm} \}$, in terms of which (34a) reads
\begin{mathletters}
\begin{equation}
{\cal L} = P_+ \dot{X}^+ + P_- \dot{X}^- + \Pi \dot{\varphi} - \lambda^+
     \left( P_+ {X^+}' + \theta_+^{\rm matter} \right) - \lambda^-
          \left(- P_- {X^-}' + \theta_-^{\rm matter} \right)
\end{equation}
\begin{equation}
\theta_{\pm}^{\rm matter} = {1 \over 4} (\Pi \pm \varphi')^2
\end{equation}
\end{mathletters}%
The gravity portions of the constraints $\Theta_{\pm}$ have been transformed to
$\pm P_{\pm} {X^{\pm}}'$ ---
expressions that look like momentum densities for fields
$X^{\pm}$, and thus satisfy the
$\Theta_\pm$ algebra (33) without
center, as do also momentum densities, see (12a).

The entire obstruction in the full gravity plus matter constraints comes
from the commutator of the matter contributions $\theta_\pm^{\rm
matter}$.  In order to remove the obstruction, we modify the theory by
adding $\Delta \Theta_\pm$ to the constraint $\Theta_\pm$, such that no
center arises in the modified constraints.  An expression for
$\Delta \Theta_\pm$ that does the job is
\begin{equation}
\Delta \Theta_{\pm} = {1 \over 48 \pi} {({\rm ln} {X^{\pm}}')}''
\end{equation}
Hence $\tilde{\Theta}_{\pm} \equiv \Theta_{\pm} + \Delta \Theta_{\pm}$
possess no obstruction in its algebra, and can annihilate states.
Explicitly, the
modified constraint equations read in the Schr\"odinger representation (after
dividing by ${X^{\pm}}'$)
\begin{equation}
\left( {1 \over i} {\delta \over \delta X^{\pm}} \pm {1 \over 48 \pi{X^{\pm}}'}
        \left({\rm ln} {X^{\pm}}'\right)'' \pm {1 \over { X^{\pm}}'}
         \theta_{\pm}^{\rm matter} \right) \psi  (X^{\pm}, \varphi) = 0
\end{equation}
It is recognized that the anomaly has been removed by introducing
functional $U(1)$ connections in $X^{\pm}$ space, whose curvature
cancels the anomaly.  In the modified constraint there still is no
mixing between gravitational variables $\{ P_{\pm}, X^{\pm} \}$ and
matter variables $\{ \Pi, \varphi \}$.  But the modified gravitational
contribution is no longer quadratic --- indeed it is non-polynomial ---
and we have no idea how to solve (37).  We suspect, however, that just as
its matter-free version, Eq.~(37) possesses only a few solutions --- far
fewer than the rich spectrum that emerges upon BRST quantization, which
we now examine.

In the BRST quantization method, extensively employed by string and
conformal field theory investigators, one adds ghosts, which carry their
own anomaly of $c_{\rm ghost} = -26$.  Also one improves $\Theta_\pm$ by
the addition of $\Delta\Theta_\pm$ so that $c$ is increased; for
example, with
\begin{eqnarray}
\Delta \Theta_\pm &=& {Q \over \sqrt{4\pi}} \left( \Pi \pm \varphi'
\right)'
\\
c &\to& c + 3 Q^2
\end{eqnarray}
[The modification (38) corresponds to ``improving''
the energy momentum tensor by
$(\partial_\mu \partial_\nu - g_{\mu\nu} \Box ) \varphi$].
The ``background charge'' $Q$ is chosen so that the total anomaly
vanishes.
\be
c + 3Q^2 + c_{\rm ghost} = 0
\ee
Moreover, the constraints are relaxed by imposing that physical states
are annihilated by the ``BRST'' charges, rather than by the bosonic
constraints.  This is roughly equivalent to enforcing ``half'' the
bosonic constraints, the positive frequency portions.  In this way one
arrives at a rich and well known spectrum.

Within BRST quantization, the negative signed gravitational field $r^0$
is quantized so that negative norm states arise --- just as in
Gupta-Bleuler electrodynamics.   [Negative norm states cannot arise in a
Schr\"odinger representation, where the inner product is explicitly
given by a (functional) integral, leading to positive norm.]~
One then finds $c_0 = 1$; the center is insensitive to the signature
with which fields enter the action.   As a consequence, $c^{\rm gravity}
= c_0 + c_1 = 2$ so that even pure gravity constraints possess an
obstruction.

Evidently, pure gravity with $c^{\rm gravity} = 2$ requires $Q = 2
\sqrt{2}$.  The rich BRST spectrum is much more plentiful than the two
states (31) found in the Schr\"odinger representation
and does not appear to reflect the fact that the classical pure gravity
theory is without excitations.

Gravity with matter carries $c=3$, and becomes quantizable at $Q =
\sqrt{23/3}$.  Once again a rich spectrum emerges, but it shows no
apparent relation to a particle spectrum.

\vspace*{.2in}
\begin{center}
{\large CONCLUSIONS}
\end{center}

Without question, the CGHS model, and other similar two-dimensional
gravity models, are afflicted by anomalies in their constraint algebras,
which become second-class and frustrate straightforward quantization.
While anomalies can be calculated and are finite, their specific value
depends on the way singularities of quantum field theory are resolved,
and this leads to a variety of procedures for overcoming the problem and
to a variety of quantum field theories, with quite different properties.

Two methods were discussed:
(i) a Schr\"odinger representation with
Kucha\v{r}-type improvement
as needed, {\it i.e.\/} when matter is present,
and (ii) BRST quantization.  (Actually several
other approaches are also available [7].)  Only in the
first method for pure matterless gravity,
with positive norm states and vanishing anomaly,
does the quantum
theory bear any resemblance to the classical theory,
in the sense that the classical gravity theory has no propagating
degrees of freedom, while the quantum Hilbert space has only
the two states in (31), neither of which contains any further degrees of
freedom.  In other cases,
{\it e.g.}~with matter, the classical picture of physics seems
irrelevant to the behavior of the quantum theory.

Presumably, if anomalies were absent, the different quantization
procedures (Schr\"odinger representation, BRST, $\ldots$) would
produce the same physics.  However, the anomalies {\it are\/} present and
interfere with equivalence.

Finally, we remark that our investigation has exposed an interesting
structure within Virasoro theory: there exists a field theoretic
realization of the algebra without the anomaly, in terms of spinless
fields and with no ghost fields.   Moreover, there are states that are
annihilated by {\it all\/} the Virasoro generators.

What does any of this teach us about the physical four-dimensional
model?  We believe that an extension in the constraint algebra will
arise for all physical, propagating degrees of freedom: for matter
fields, as is seen already in two dimensions, and also for gravity
fields, which in four dimensions (unlike in two) carry physical energy.
How to overcome this obstruction to quantization is unclear to us, but
we expect that the resulting quantum theory will be far different from
its classical counterpart.  Especially problematic is the fact that
flat-space calculations of anomalous Schwinger terms in four dimensions
yield infinite results, essentially for dimensional reasons.
Moreover, it should be clear that any
announced ``solutions'' to the constraints that result from
{\it formal\/} analysis must be viewed as preliminary:
properties of the Hilbert space and of the inner
product must be fixed first in order to give an unambiguous
determination of any obstructions.

We believe that our two-dimensional investigation, although in a much simpler
and unphysical setting, nevertheless contains important clues for
realistic theories.  Certainly that was the lesson of gauge theories:
anomalies and vacuum angle have corresponding roles in the Schwinger
model and in QCD!


\begin{thebibliography}{99}

\bibitem{1}
See R.~Jackiw, in
{\it Quantum Mechanics of Fundamental Systems\/},
C.~Teitelboim, ed.
(Plenum, New York NY, 1988).

\bibitem{2}
D.~Boulware and S.~Deser,
{\it J.~Math.~Phys\/}~{\bf 8}, 1468 (1967).

\bibitem{3}
R.~Jackiw, C.~Teitelboim, in
{\it Quantum Theory of Gravity\/},
S.~Christensen, ed.
(A.~Hilger, Bristol UK, 1984).

\bibitem{4}
D.~Cangemi,
{\it Phys.~Lett.\/}~{\bf B297}, 261 (1992);
A.~Ach\'ucarro,
{\it Phys.~Rev.~Lett.\/}~{\bf 70}, 1037 (1993).

\bibitem{5}
A.~Polyakov,
{\it Mod.~Phys.~Lett.\/}~{\bf A2}, 893 (1987),
{\it Gauge Fields and Strings\/}
(Harwood, New York NY, 1987).

\bibitem{6}
C.~Callan, S.~Giddings, J.~Harvey and A.~Strominger,
{\it Phys.~Rev.~D\/}~{\bf 45} 1005 (1992);
H.~Verlinde in
{\it Sixth Marcel Grossmann Meeting on General Relativity\/},
M.~Sato and T.~Nakamura, eds.
(World Scientific, Singapore, 1992).

\bibitem{7}
Our most recent paper on this entire subject, with reference to earlier
work, is D.~Cangemi, R.~Jackiw and B.~Zwiebach,
{\it Ann.~Phys.\/} (NY) (in press).

\bibitem{8}
P.~Jordan, {\it Z.~Phys.\/}~{\bf 93}, 464 (1935).
In the context of the energy-momentum tensor anomaly, this is explained
in R.~Floreanini and R.~Jackiw,
{\it Phys.~Lett.\/}~{\bf B175}, 428 (1986).

\bibitem{9}
D.~Cangemi and R.~Jackiw,
{\it Phys.~Lett.\/}~{\bf B337}, 271 (1994),
{\it Phys.~Rev.~D\/}~{\bf 50}, 3913 (1994);
D.~Amati, S.~Elitzur and E.~Rabinovici,
{\it Nucl.~Phys.\/}~{\bf B418}, 45 (1994);
D.~Louis-Martinez, J.~Gegenberg and G.~Kunstatter,
{\it Phys.~Lett.\/}~{\bf B321}, 193 (1994);
E.~Benedict,
{\it Phys.~Lett.\/}~{\bf B340}, 43 (1994);
T.~Strobl,
{\it Phys.~Rev.~D\/}~{\bf 50}, 7346 (1994).

\bibitem{10}
K.~Kucha\v{r}, {\it Phys.~Rev.~D\/}~{\bf 39}, 2263 (1989);
K.~Kucha\v{r} and G.~Torre, {\it J.~Math.~Phys.\/}~{\bf 30}, 1769 (1989).

\end{thebibliography}
\end{document}